\documentclass[twocolumn]{svjour3}          
\pdfoutput=1

\usepackage{graphicx}
\usepackage{xspace}
\usepackage{textcomp}
\usepackage{cite}
\usepackage[T1]{fontenc}
\usepackage{graphicx}
\usepackage{amsmath}
\usepackage{amssymb}
\usepackage[binary,textstyle,amssymb,thinqspace]{SIunits}
\usepackage{multirow}
\usepackage{rotating}
\usepackage{setspace}
\usepackage{booktabs}
\usepackage{xspace}
\usepackage{algorithm}
\usepackage{algpseudocode}
\usepackage{stfloats}
\usepackage[toc,nonumberlist]{glossaries}
\usepackage{import}
\usepackage{array}
\usepackage[numbers]{natbib}
\usepackage{color}
\usepackage{caption}

\renewcommand{\eqref}[1]{(\ref{#1})}    

\newcommand{\secref}[1]{\mbox{Section~\ref{#1}}}
\newcommand{\figref}[1]{\mbox{Fig.~\ref{#1}}}
\newcommand{\tblref}[1]{\mbox{Tbl.~\ref{#1}}}

\newcommand{\eqnref}[1]{\mbox{(\ref{#1})}}

\definecolor{rvn1}{rgb}{0.0117,0.0312,0.6875} 
\definecolor{rvn2}{rgb}{0.1484,0.8438,0.2227} 
\newcommand{\rvO}[1]{{#1}}    
\newcommand{\rvT}[1]{{#1}}    

\newcommand{\safemath}[2]{\newcommand{#1}{\ensuremath{#2}\xspace}}

\safemath{\setA}{\mathcal{A}}
\safemath{\setB}{\mathcal{B}}
\safemath{\setC}{\mathcal{C}}
\safemath{\setD}{\mathcal{D}}
\safemath{\setE}{\mathcal{E}}
\safemath{\setF}{\mathcal{F}}
\safemath{\setG}{\mathcal{G}}
\safemath{\setH}{\mathcal{H}}
\safemath{\setI}{\mathcal{I}}
\safemath{\setJ}{\mathcal{J}}
\safemath{\setK}{\mathcal{K}}
\safemath{\setL}{\mathcal{L}}
\safemath{\setM}{\mathcal{M}}
\safemath{\setN}{\mathcal{N}}
\safemath{\setO}{\mathcal{O}}
\safemath{\setP}{\mathcal{P}}
\safemath{\setQ}{\mathcal{Q}}
\safemath{\setR}{\mathcal{R}}
\safemath{\setS}{\mathcal{S}}
\safemath{\setT}{\mathcal{T}}
\safemath{\setU}{\mathcal{U}}
\safemath{\setV}{\mathcal{V}}
\safemath{\setW}{\mathcal{W}}
\safemath{\setX}{\mathcal{X}}
\safemath{\setY}{\mathcal{Y}}
\safemath{\setZ}{\mathcal{Z}}

\safemath{\bma}{\mathbf{a}}
\safemath{\bmb}{\mathbf{b}}
\safemath{\bmc}{\mathbf{c}}
\safemath{\bmd}{\mathbf{d}}
\safemath{\bme}{\mathbf{e}}
\safemath{\bmf}{\mathbf{f}}
\safemath{\bmg}{\mathbf{g}}
\safemath{\bmh}{\mathbf{h}}
\safemath{\bmi}{\mathbf{i}}
\safemath{\bmj}{\mathbf{j}}
\safemath{\bmk}{\mathbf{k}}
\safemath{\bml}{\mathbf{l}}
\safemath{\bmm}{\mathbf{m}}
\safemath{\bmn}{\mathbf{n}}
\safemath{\bmo}{\mathbf{o}}
\safemath{\bmp}{\mathbf{p}}
\safemath{\bmq}{\mathbf{q}}
\safemath{\bmr}{\mathbf{r}}
\safemath{\bms}{\mathbf{s}}
\safemath{\bmt}{\mathbf{t}}
\safemath{\bmu}{\mathbf{u}}
\safemath{\bmv}{\mathbf{v}}
\safemath{\bmw}{\mathbf{w}}
\safemath{\bmx}{\mathbf{x}}
\safemath{\bmy}{\mathbf{y}}
\safemath{\bmz}{\mathbf{z}}

\safemath{\bA}{\mathbf{A}}
\safemath{\bB}{\mathbf{B}}
\safemath{\bC}{\mathbf{C}}
\safemath{\bD}{\mathbf{D}}
\safemath{\bE}{\mathbf{E}}
\safemath{\bF}{\mathbf{F}}
\safemath{\bG}{\mathbf{G}}
\safemath{\bH}{\mathbf{H}}
\safemath{\bI}{\mathbf{I}}
\safemath{\bJ}{\mathbf{J}}
\safemath{\bK}{\mathbf{K}}
\safemath{\bL}{\mathbf{L}}
\safemath{\bM}{\mathbf{M}}
\safemath{\bN}{\mathbf{N}}
\safemath{\bO}{\mathbf{O}}
\safemath{\bP}{\mathbf{P}}
\safemath{\bQ}{\mathbf{Q}}
\safemath{\bR}{\mathbf{R}}
\safemath{\bS}{\mathbf{S}}
\safemath{\bT}{\mathbf{T}}
\safemath{\bU}{\mathbf{U}}
\safemath{\bV}{\mathbf{V}}
\safemath{\bW}{\mathbf{W}}
\safemath{\bX}{\mathbf{X}}
\safemath{\bY}{\mathbf{Y}}
\safemath{\bZ}{\mathbf{Z}}
\safemath{\bZero}{\mathbf{0}}
\safemath{\bPsi}{\mathbf{\Psi}}
\safemath{\bTheta}{\mathbf{\Theta}}
\safemath{\bDelta}{\mathbf{\Delta}}

\newcommand{\NES}{\ensuremath{\mathrm{N}_{es}}\xspace}
\newcommand{\NSD}{\ensuremath{\mathrm{N}_{sd}}\xspace}
\newcommand{\NSTS}{\ensuremath{\mathrm{N}_{sts}}\xspace}
\newcommand{\NPE}{\ensuremath{\mathrm{N}_{pe}}\xspace}

\newcommand{\NTX}{$\mathrm{N}_{TX}$}
\newcommand{\NRX}{$\mathrm{N}_{RX}$}
\newcommand{\NSS}{$\mathrm{N}_{SS}$}

\def\addnotation #1: #2{\parbox[t]{1.25cm}{$#1$\dotfill} \parbox[t]{3.50in}{#2} \vspace{0.0625cm} \\}
\def\addsymbol #1: #2{\parbox[t]{2cm}{$#1$} \parbox[t]{3in}{#2} \vspace{0.0625cm} \\}
\def\addacronym #1: #2{\parbox[t]{2cm}{\bf{#1}\dotfill} \parbox[t]{2.8in}{#2} \vspace{0.0625cm} \\}

\DeclareMathOperator*{\argmin}{arg\;min}
\DeclareMathOperator*{\argmax}{arg\;max}

\newacronym{MIMO}{MIMO}{multiple-input multiple-output}
\newacronym{LD}{LD}{linear detector}
\newacronym{LRALD}{LRALD}{lattice reduction aided linear detector}
\newacronym{OFDM}{OFDM}{orthogonal frequency division multiplexing}
\newacronym{SISO}{SISO}{single-input single-output}
\newacronym{COSIC}{COSIC}{conditioned ordered successive interference cancellation}
\newacronym{STS}{STS}{single tree search sphere decoder}
\newacronym{PHY}{PHY}{physical}
\newacronym{ASIC}{ASIC}{application specific integrated circuit}
\newacronym{LR}{LR}{lattice reduction}
\newacronym{MAC}{MAC}{medium access control}
\newacronym{OSI}{OSI}{open systems interconnection}
\newacronym{HT}{HT}{high throughput}
\newacronym{non-HT}{non-HT}{non high throughput}
\newacronym{GF}{GF}{\gls{HT} green field}
\newacronym{MF}{MF}{\gls{HT}-mixed format}
\newacronym{GI}{GI}{guard interval}
\newacronym{AHB}{AHB}{advanced high-performance bus}
\newacronym{AMBA}{AMBA}{advanced microcontroller bus architecture}
\newacronym{MCS}{MCS}{modulation and coding scheme}
\newacronym{FFT}{FFT}{fast Fourier transform}
\newacronym{IFFT}{IFFT}{inverse fast Fourier transform}
\newacronym{CSI}{CSI}{channel state information}
\newacronym{RTS}{RTS}{ready-to-send}
\newacronym{CTS}{CTS}{clear-to-send}
\newacronym{ACK}{ACK}{acknowledgments}
\newacronym{HR}{HR}{heart rate}
\newacronym{FPGA}{FPGA}{fieldprogrammable logic array}
\newacronym{FSM}{FSM}{finite state machine}
\newacronym{SVD}{SVD}{singular value decomposition}
\newacronym{AT}{AT}{area times time}
\newacronym{MGS}{MGS}{modified Gram-Schmidt}
\newacronym{LLR}{LLR}{log-likelihood ratio}
\newacronym{SNR}{SNR}{signal to noise ratio}
\newacronym{SINR}{SINR}{signal to interference and noise ratio}
\newacronym{flops}{flops}{floating point operations}
\newacronym{CORDIC}{CORDIC}{coordinate rotation digital computer}
\newacronym{LUT}{LUT}{look-up table}
\newacronym{SIC}{SIC}{successive interference cancellation}
\newacronym{ML}{ML}{maximum likelihood}
\newacronym{APP}{APP}{a posteriori probability}
\newacronym{DAC}{DAC}{digital to analog converter}
\newacronym{BFP}{BFP}{block-floating-point}
\newacronym{TGn}{TGn}{task-group-n}
\newacronym{VLSI}{VLSI}{very large scaled integrated}
\newacronym{DVFS}{DVFS}{dynamic voltage and frequency scaling}
\newacronym{DAB}{DAB}{digital audio broadcasting}
\newacronym{PAPR}{PAPR}{peak-to-average power ratio}
\newacronym{CMOS}{CMOS}{complementary metal-oxide-semiconductor}
\newacronym{PAM}{PAM}{pulse amplitude modulation}
\newacronym{DVB}{DVB}{digital video broadcast}
\newacronym{ISI}{ISI}{inter-symbol interference}
\newacronym{MMSE}{MMSE}{minimum mean squared error}
\newacronym{ZF}{ZF}{zero forcing}
\newacronym{ASIP}{ASIP}{application specific instruction-set processor}
\newacronym{MAP}{MAP}{maximum a posteriori}
\newacronym{AGC}{AGC}{automated gain control}
\newacronym{HDL}{HDL}{hardware description language}
\newacronym{PED}{PED}{partial Euclidean distance}
\newacronym{ED}{ED}{Euclidean distance}
\newacronym{SA}{SA}{Seysen's algorithm}
\newacronym{SD}{SD}{sphere decoder}
\newacronym{NR}{NR}{Newton-Raphson}
\newacronym{BER}{BER}{bit error rate}
\newacronym{FER}{FER}{frame error rate}
\newacronym{LLL}{LLL}{Lenstra-Lenstra-Lov\'asz}
\newacronym{FEC}{FEC}{forward error correction}
\newacronym{BCC}{BCC}{binary convolutional code}
\newacronym{LDPC}{LDPC}{low density parity check}
\newacronym{STBC}{STBC}{space time block code}
\newacronym{CSD}{CSD}{cyclic shift}
\newacronym{STF}{STF}{short training field}
\newacronym{LTF}{LTF}{long training field}
\newacronym{SIG}{SIG}{signal}
\newacronym{CRC}{CRC}{cyclic redundancy check}
\newacronym{IFS}{IFS}{inter-frame spacing}
\newacronym{SIFS}{SIFS}{short \gls{IFS}}
\newacronym{PLCP}{PLCP}{physical layer convergence protocol} 
\newacronym{PMD}{PMD}{physical medium dependent} 
\newacronym{PPDU}{PPDU}{PLCP protocol data unit} 
\newacronym{PSDU}{PSDU}{PLCP service data unit}  
\newacronym{MPDU}{MPDU}{MAC protocol data units} 
\newacronym{PLME}{PLME}{physical layer management entity} 
\newacronym{VGA}{VGA}{variable gain amplifier}
\newacronym{ADC}{ADC}{analog-to-digital converter}
\newacronym{CFO}{CFO}{carrier-frequency offset}
\newacronym{SRO}{SRO}{sampling rate offset}
\newacronym{CP}{CP}{cyclic prefix}
\newacronym{LS}{LS}{least square}
\newacronym{PE}{PE}{processing element}
\newacronym{NSS}{\NSS}{number of spatial streams}
\newacronym{NSTS}{\NSTS}{number of space-time streams}
\newacronym{NES}{\NES}{number of encoded streams}
\newacronym{NSD}{\NSD}{number of OFDM tones per OFDM symbol}
\newacronym{NTX}{\NTX}{number of transmit antennas}
\newacronym{NRX}{\NRX}{number of receive antennas}
\newacronym{NPE}{\NPE}{number of \glspl{PE}}
\newacronym{PLL}{PLL}{phase locked loop}
\newacronym{FO}{FO}{frequency offset}
\newacronym{LNA}{LNA}{low noise amplifier}
\newacronym{AP}{AP}{access point}
\newacronym{VCD}{VCD}{value change dump}
\newacronym{TDD}{TDD}{time-division-duplex}
\newacronym{TDMA}{TDMA}{time-division-multiple-access}
\newacronym{RA}{RA}{rate adaptation}
\newacronym{WLAN}{WLAN}{wireless local area network}
\newacronym{GE}{GE}{Goodput-Aware Energy-Guided}
\newacronym{EG}{EG}{Energy-Guided}
\newacronym{GG}{GG}{Goodput-Guided}
\newacronym{NIC}{NIC}{network interface card}
\newacronym{IC}{IC}{integrated circuit}

\begin{document}


\title{Cross-layer Energy-Efficiency Optimization of Packet Based Wireless MIMO Communication Systems
}


\author{Christian Senning         \and
        Georgios Karakonstantis   \and
        Andreas Burg
}


\institute{Christian Senning, Georgios Karakonstantis Andreas Burg \at
           EPFL-STI-IEL-TCL, Station 11, ELG 011,  \\
           Tel.: +41 21 693 6924\\
           Fax: +41 21 693 2687\\
           \email{\{christian.senning, georgios.karakonstantis, andreas.burg@epfl.ch\}}           \\
           This paper extends the work published in \\ICASSP \citep{senning14ICASSP}
}

\date{Received: date / Accepted: date}

\maketitle

\begin{abstract}
Energy in today's short-range wireless communication is mostly spent on the analog- and digital hardware rather than on radiated power. Hence, purely information-theoretic considerations fail to achieve the lowest energy per information bit and the optimization process must carefully consider the overall transceiver. In this paper, we propose to perform cross-layer optimization, based on an energy-aware rate adaptation scheme combined with a physical layer that is able to properly adjust its processing effort to the data rate and the channel conditions to minimize the energy consumption per information bit. This energy proportional behavior is enabled by extending the classical system modes with additional configuration parameters at the various layers. Fine grained models of the power consumption of the hardware are developed to provide awareness of the physical layer capabilities to the medium access control layer. 
\rvT{The joint application of the proposed energy-aware rate adaptation and modifications to the physical layer of an IEEE\,802.11n system, improves energy-efficiency (averaged over many noise and channel realizations) in all considered scenarios by up to 44\%.}

\end{abstract}

\section{Introduction}\label{sec:intro}
Mobile communication -- anytime, anywhere access to data and communication services -- has been continuously increasing since the operation of the first cellular phone system. 
This growth is combined with an increasing demand by consumers for small and multifunctional products. Such products must be able to transmit wirelessly not only voice, but also color pictures and video, as well as to provide access to complex applications over the worldwide web \citep{man05}. 
Satisfying such a growing demand for wireless communications has been achieved by advances in information theory, combined with the ability to manufacture high throughput communication circuits. Unfortunately, concerns are being raised that this scenario cannot continue forever. 
Specifically, as more and more transistors are packed onto a single chip to support the associated demand for high performance communications, more of them toggle in ``active'' mode or leak during ``sleep'' periods, resulting in a substantial increase of on-chip power dissipation~\citep{li11}. 
Such increased power dissipation in combination with the slow improvements in battery capacity limit the battery run-time of portable devices and prevent them from adequately meeting user expectations. 

The classical optimization for high-performance in terms of throughput or error rate \citep{Jensen10,Martorell11} performed by goodput-guided \gls{RA} schemes may not be able to close the gap between supplied and required energy. This problem may be attributed to the fact that either the best error-rate might not always be needed, or the highest throughput might exceed application requirements. In particular the peak data rates in the widely used IEEE\,802.11n standard (600\,Mbps) significantly surmount the requirements for high-quality audio transmissions (192\,kbps) or high-definition video-on-demand services (e.g., YouTube.com limits the maximum rate of its content to 5\,Mbps). This means that we have reached a point where the peak data rates of the wireless communication system is no longer always a limiting factor. This can be argued especially for scenarios where the cumulative data rate requested by all users in the same \gls{WLAN} channel is far less than the system capacity, as in isolated networks with small number of users (e.g., networks within detached houses or small offices). 
Therefore, we show that there are many opportunities which can be exploited by alternative \gls{RA} schemes to enhance the battery run-time of portable devices.

Some works have proposed energy-efficient RA schemes by mainly focusing on the minimization of the radiated power only. However, such optimization may not be so beneficial from an energy-efficiency point of view for today's networks for short range wireless connectivity, which usually spend most of the power in the signal processing of the receiver~\citep{halperin10}. Recent studies shift the throughput objective of RA schemes to energy-efficiency optimization of the overall system \citep{li09, halperin10}. Although, such works improved the energy-efficiency of wireless systems, they have not yet revealed all the potential gains, since i)~either they focus solely on the RA, neglecting possible tuning knobs on other layers or ii)~their knowledge of the \gls{PHY} layer is based on high-level information theoretical energy models, ignoring the inter dependencies across all layers and modules or iii)~they investigate a given, fixed PHY layer treated as black-box ignoring the potential to adjust processing energy according to data rate or channel conditions (energy proportionality).

\textbf{\textit{Contributions:}}
In this paper, we propose to enhance the energy-efficiency by a
cross-layer approach,  based on  energy-aware RA schemes combined with an \textit{energy proportional} PHY layer design i.e., a receiver that is able to scale its effort (i.e., its energy consumption) according to the varying complexity to  successfully recover the transmitted bits under all given operating conditions. Specifically, our contributions can be summarized as follows:
\begin{itemize}
\item An energy-guided RA scheme is developed and the achievable gains in terms of energy spend per successfully received information bit are analyzed. Our results indicate that such a RA scheme can result in significant improvements in energy-efficiency, while still satisfying the data rate of most today's applications.
\item A compromise between the energy-guided RA and the classical goodput-guided RA is presented for applications with high data rate requirements.
\item Both proposed RA schemes are enabled through a fine-grained energy model of the PHY layer that allows to capture many inter-dependencies across various settings throughout the operation of a packet based wireless system.
\item The available choices of classical RA, assumed by most existing works, are extended by exploiting circuit level techniques combined with algorithmic effort scaling. This allows to truly realize an energy proportional PHY layer and maximize the energy-efficiency gains compared to the classical throughput-guided RA that does not exploit any of the proposed PHY layer modifications.
\item A case study based on an IEEE\,802.11n compliant PHY is shown for fixed and varying packet lengths.
\end{itemize}

\textbf{\textit{Outline:}}
\secref{sec:prior} discusses prior publications for energy-efficient wireless communication systems and corresponding energy models. \secref{sec:prelim} elaborates on the background regarding a typical frame-based communication system and \secref{sec:propapp} presents the proposed approach. The required energy model is developed in \secref{subsec:em}, while the applied techniques for enabling an energy proportional PHY layer are described in \secref{sec:phyMod}. The proposed optimizations are applied to a typical IEEE 802.11n environment and the results are presented in \secref{sec:simRes}. Finally, conclusions are drawn in \secref{sec:conc}.

\section{State of the Art}\label{sec:prior}
\rvO{The traditional research has been focusing on optimizing the transmit power based on} information-theoretic energy-efficiency metrics \citep{cui04,kim09}. 
However, the proposed solutions start from a model in which the device power consumption is essentially given by the radiated energy \citep{Schurgers01,Lin06}. 
Unfortunately, this assumption rarely holds in practice as for most battery-operated systems used for wireless connectivity (e.g., WLAN) the radiated power is only a small part of its total power consumption.
Later studies have recognized this fact and have started to take active power consumed by the hardware into account. 
While this defies the straightforward application of information theoretic tools and principles, it also provides more opportunities for energy savings~\citep{Zompakis13}. 

\rvO{The hardware power has been accounted for in some early studies, e.g., \citep{app01}}.
However, the point-to-point streaming-system under consideration in \citep{app01} that only considers a physical layer, the corresponding traffic pattern, and the associated tuning knobs has little in common with today's omnipresent short-range wireless networks (e.g., IEEE\,802.11n or IEEE\,802.11ac). 
Based on this observation recent works \rvO{have been focusing} on providing strategies to improve energy-efficiency at the medium-access-control (MAC) layer \citep{pol08,tsa11}. 
The latest studies on the component level consider different algorithm choices under the aspect of energy-efficiency, spanning both the algorithm and the architectural layer \citep{kie11}.

\rvO{Later publications introduced} the concept of energy-aware baseband processing, where scalability is put to service to adjust the processing to the dynamically changing environment, leading to even better overall energy-efficiency \citep{li09,Kim08,Kim10}. 
In this context, a recent work by Intel tried to provide realistic power measurements of the baseband for the popular IEEE\,802.11n standard claiming to be the first ever doing it, indicating a shift in design objectives from throughput oriented communications to energy-efficiency oriented communication \citep{halperin10}. 
In particular, \citep{Zompakis13} showed that the baseband processing of software-defined radios is most of the time underutilized and states that this underutilization can be used for power reduction (resulting in the desired energy proportional behavior). 

In order to exploit such underutilization, the MAC layer has to be  aware of the energy consumption of its associated PHY layer.
For this purpose, many models for the energy consumption of wireless systems have been proposed in the literature. Some models focus exclusively on the transmit power~\citep{kim09} and do not capture the main energy-drains, while others jointly model the baseband processing of the transmitter and the receiver \citep{Auer11,Thandapani12}. These models then also result in optimization of the grid-powered access points instead of focusing on the more critical battery operated devices.
Further works rely on measurements of  entire chip sets or network interface cards \citep{halperin10,Li12,Tauber12,carroll10} that do not provide insight into the energy consumption of the PHY layer itself. 

Although existing studies have indicated the need for improving energy-efficiency, there is still a need for true cross-layer optimization, enabled by an energy-guided RA that is aware of the energy consumption of an associated energy proportional PHY layer, through accurate models.

\begin{figure}
 \centering
 \includegraphics[width=0.68\columnwidth]{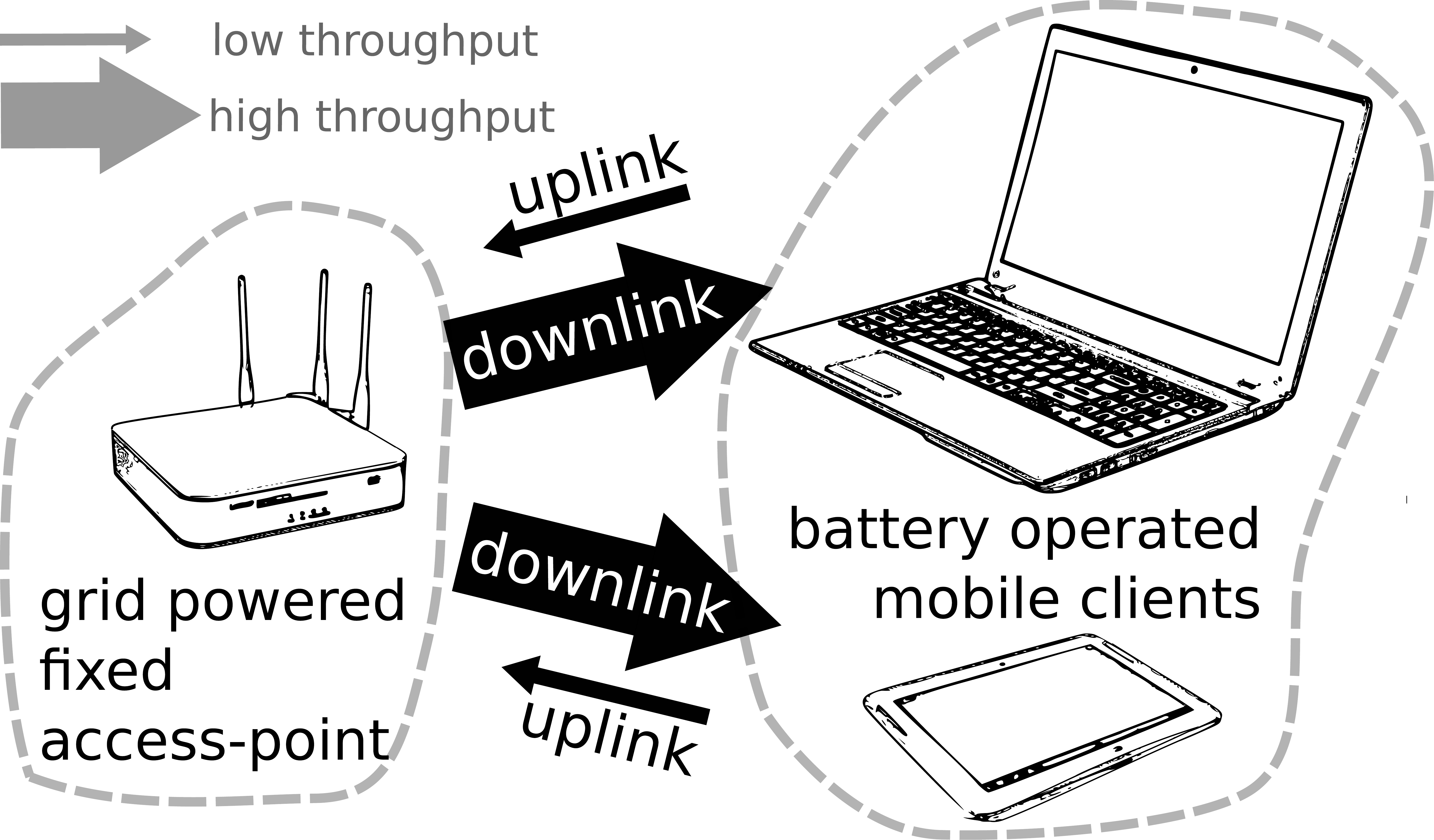}
 \caption{WLAN setup with high traffic load from access point to battery operated mobile clients.}
 \label{ch8:fig:comsys}
\end{figure}

\section{Background} \label{sec:prelim}
\figref{ch8:fig:comsys} illustrates a point-to-point link of a frame-based wireless communication system, which is typical for short-range \gls{WLAN} networks. \rvO{The essential components for our consideration are a grid powered, central \gls{AP} and one or more battery operated devices that are referred to as \emph{clients}. In our scenario, all clients receive application data (e.g., audio or video) transmitted from the AP over a wireless channel.} 

Both the AP and the client, capable of transmitting and receiving multiple data streams using multiple antennas, consist of a PHY layer that is responsible for the actual transmission and reception of the data as well as a MAC layer that controls the PHY layer. The PHY layer is composed of various hardware modules. In transmit mode, these modules encode the application data, modulate them, convert them into analog radio frequency (RF) signals, and finally radiate them over the available antenna(s). On the other side of the point-to-point link, the PHY layer operates in receive mode and its modules apply the opposite functions converting the received RF signals into digital binary data, demodulating and decoding them as well as finally extracting the application data. 
 
We assume a system that operates in \gls{TDD} and \gls{TDMA} mode and we consider a typical asymmetric scenario in which data is mostly downloaded from the AP to the client. Hence, the client is most of the time in receive mode. However, there are also some short time-slots during which each client gets into transmit mode for sending acknowledgment (ACK) frames to the AP. 

\subsection{Packet Structure}
\begin{figure}[t]
\centering
\includegraphics[width=1.0\columnwidth]{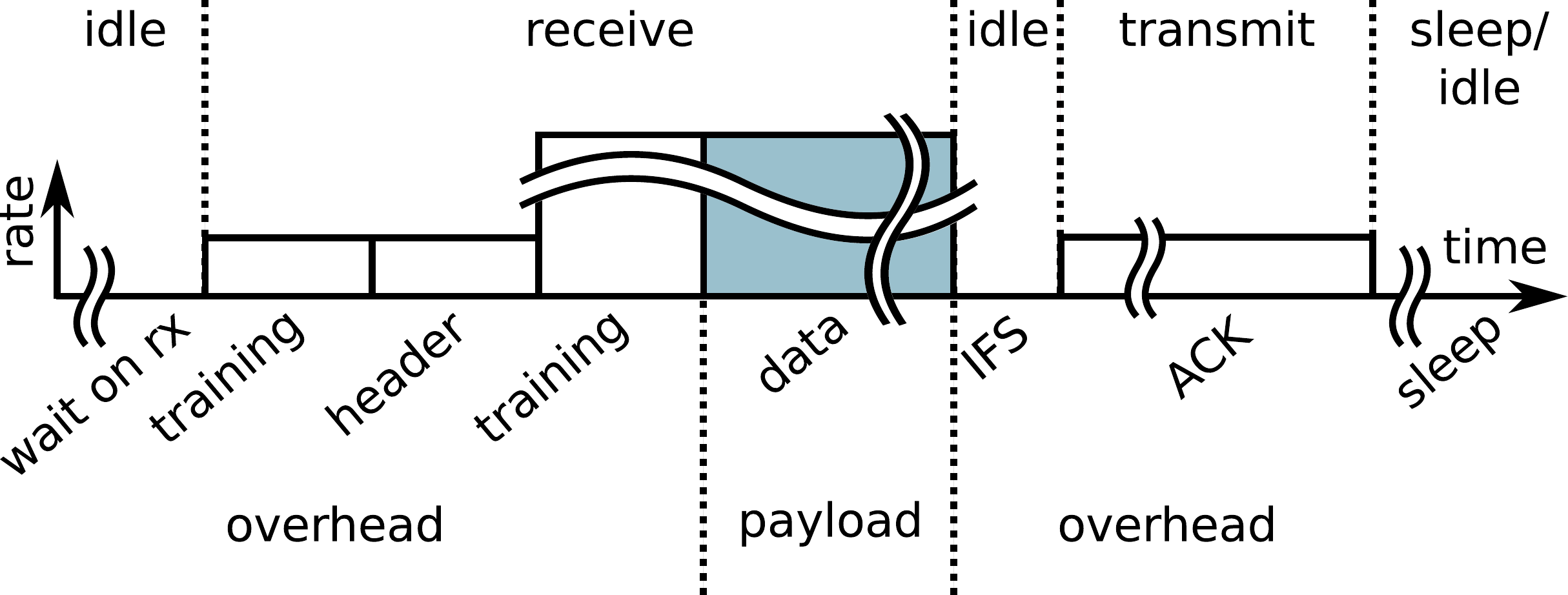}
\caption{Simplified illustration of the IEEE\,802.11n transmission protocol.}
\label{ch8:fig:frame}
\end{figure}

As illustrated in \figref{ch8:fig:frame}, the transmission sequence begins with a frame start waiting period. The duration of this period depends on proper sleep time prediction based on the power save poll MAC protocol. The actual frame starts with a training sequence used for frame start detection, initial frequency offset estimation, and channel estimation. The initial channel estimate is then used to detect the frame header containing information about the subsequent data payload (e.g., the payload length in number of bytes and the \gls{MCS}, which together determine the frame length in number of symbols). If the number of spatial streams differs between header and payload, an additional training sequence is required to enable the estimation of the \gls{MIMO} channel required later for detection of the data payload. The final part of the received frame is the data payload itself. Within an inter-frame spacing (IFS), the receiver has to decide if the data payload was correctly extracted and has to start transmitting an ACK frame. After sending the ACK frame, the PHY layer goes into sleep mode until he is again triggered by the MAC layer for the next frame reception.

\subsection{Classical Rate Adaptation}

A key feature of most communication systems for short-range wireless connectivity, as described above, is their ability to support a variety of transmission schemes to adjust the rate to the current channel conditions. For instance, the modulation and code rate can be selected along with other parameters such as the number of spatial streams \NSS, which together define the MCS, encoded in the header field of a packet. The transmission mode and the values of the associated parameters are selected by the MAC layer of the transmitter, based on channel state information\rvT{\footnote{\rvT{While data reception relies on accurate up-to-date channel state information, we have observed that channel characteristics relevant for RA remain stable over a long time.}}}. In today's systems, the main objective of commercially available goodput-guided (GG) RA schemes is the selection of the most appropriate system mode $\nu$ under given channel conditions for the maximization of the goodput. The system mode is selected from a set $\Omega_{GG}$ that contains all possible transmission modes. Such transmission modes determine the employed MCS and the frame length $L$ in bits. The objective of GG RA is therefore given as follows
\begin{equation}
\nu_{GG} = \argmax_{\nu} \big\{ \left(1-P_{e}(\nu)\right)\Phi(\nu) \big\},
\label{eqn:goodput}
\end{equation}
where $\Phi(\nu)$ is the throughput and $P_e(\nu)$ is the probability of a packet error. The GG RA therefore requires to estimate the error rate at run-time based on the current channel conditions. For this, a variety of known \rvO{techniques} for accurately estimating $P_{e}(\nu)$ can be applied as discussed for example in~\citep{Kim10}.

The choice of $\nu$ under such a GG RA actually determines the operating mode of both the transmitter and receiver.
Due to differences in the (de-)modula\-tion and \mbox{(de-)coding} process as well as differences in the packet structure and duration this can also lead to differences in the required signal processing effort. Hence, the MCS selection by the RA clearly has an impact on system power consumption on both sides of the wireless link. 

\begin{figure}[t]
\center
\includegraphics[width=\columnwidth]{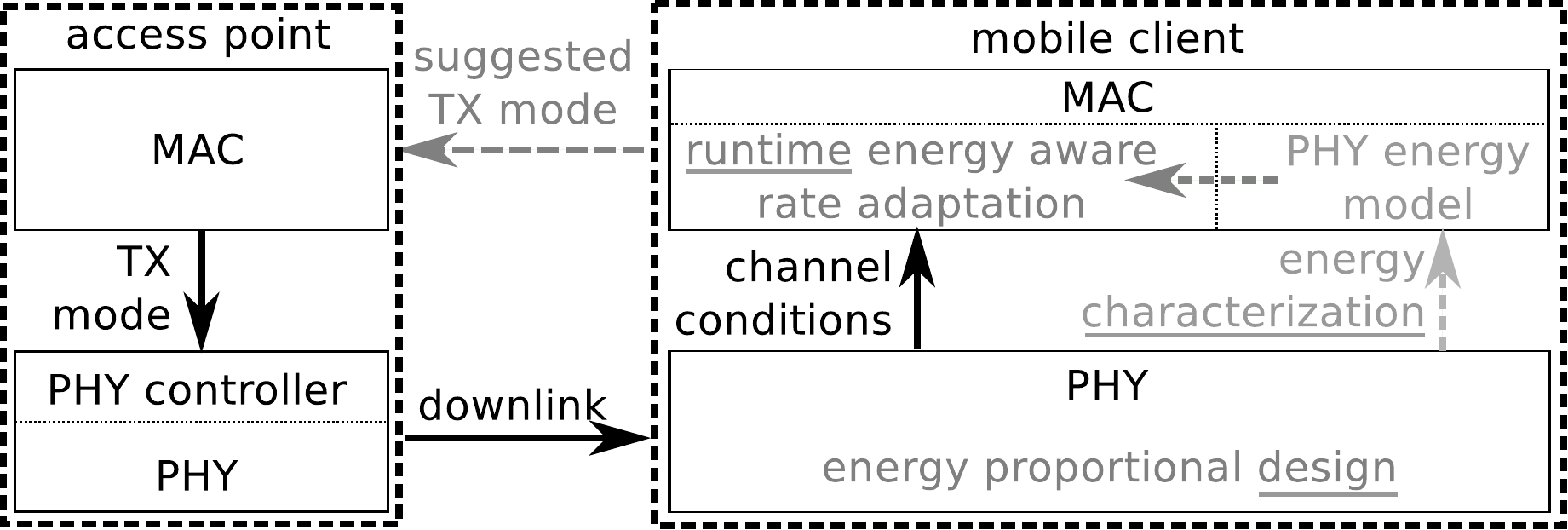}
\caption{Proposed approach for improved energy-efficiency of the
  mobile client.}
\label{ch8:fig:approach}
\end{figure}

\section{Proposed Cross-Layer Approach for Improved Energy-Efficiency}\label{sec:propapp}

As we discussed, the client in modern wireless links is in many cases a battery operated device that operates most of the time in receive mode. The amount of data transmitted to the client is usually independent of the available peak data rates but given by use patterns and applications. Therefore, the main goal of our cross-layer energy-efficiency optimization is to reduce the energy consumption per successfully received information bit at the receiver to maximize the number of bits that can be received on a single battery charge.

An illustration of the overall proposed approach is shown in \figref{ch8:fig:approach}. \rvO{The \gls{MAC} layer at the AP configures its PHY layer and forwards data to be transmitted. A packet containing this data is then sent over a wireless channel to the mobile client. The PHY layer of the client tries to recover the transmitted data based on the received signal. The energy consumption of the PHY layer of the client thereby depends on the transmission mode and the ``difficulty'' to recover the data under given channel conditions. To account for this dependency when choosing a mode, the energy consumption for different configurations are modeled in the MAC layer of the client. Based on a prediction of this energy model as well as on channel conditions reported from the PHY layer, a suitable transmission mode and corresponding receiver configurations are  suggested to the AP for future transmissions. Such a RA is discussed in the next section of this publication.}

\subsection{Energy-Guided Rate Adaptation}
In our approach, we depart from the classical GG RA and propose an energy-guided (EG) RA which aims at reducing the energy consumption of the receiver. Specifically, the method selects the system mode according to
\begin{align} \label{eqn:EG}
 \nu_{EG} = \argmin_{\nu} \left\{\eta(\nu) \right\},
\end{align}
with $\nu \in \Omega_{EG}$, where $\Omega_{EG}$ corresponds to the set of all meaningful system modes that include the modes in $\Omega_{GG}$ as well as receiver specific settings. The function $\eta(\nu)$ represents the energy consumption of the battery operated client per received information bit. Note that, at this point no attention is specifically paid towards the impact on goodput or error rate, except for the implicit dependency of $\eta(\nu)$ on the data rate $\Phi(\nu)$ and the probability of a packet error $P_e(\nu)$. As we will elaborate on, this dependency still provides a small bias toward the higher rate modes, but also respects the availability of modes associated with lower rates for the benefit of energy-efficiency. 

In contrast to the classical RA performed at the transmitter (based on channel quality feedback), energy-guided RA has to be performed at the battery operated client. This is because the AP can usually not accurately estimate or model the power consumption of a third-party battery operated device. To this end, we assume a setup in which MCSs are suggested from the client to the MAC layer of the AP\footnote{Important standards such as IEEE 802.11n already include this possibility} for the next data frame (assuming a reasonably static channel), during the ACK frames.

For occasions, when the goodput achieved by the EG RA does not satisfy the required data rate of an application, we propose a goodput-aware energy-guided (GAEG) RA that provides a compromise between the goodput achieved by the GG RA and the energy-efficiency provided by the EG RA. \rvT{Such a compromise is achieved by choosing the mode $\nu_{GAEG}$ with the best goodput and an energy-efficiency that is upper bounded by the factor $k>1$ times the energy per bit of $\nu_{EG}$ based on} 
\rvO{
\begin{equation}\label{eqn:gaeg}
\resizebox{0.91\columnwidth}{!}{$\nu_{GAEG} = \argmax_{\nu} \big\{ (1-P_e(\nu))\Phi(\nu)\ |\ \eta(\nu) < k \eta(\nu_{EG}) \big\}$}.
\end{equation}
}
\rvT{In this approach the MAC layer is able to trade energy consumption versus throughput at run time and therefore allows adjustment to the application requirements or for systems operated close to the capacity bound, by selecting an appropriate value of the factor $k$.}

\subsection{Augmenting System Modes}
In addition to shifting the RA to the client side, the proposed RA in \eqnref{eqn:EG} does not only select an MCS and the packet length $L$, but also other configuration parameters of the client. 
Such an \emph{extended set} $\Omega_{GG}$ of configuration parameters may include: i) algorithmic choices, such as different MIMO detection algorithms, ii) number of active receive chains, and iii) applied circuit level techniques, like clock-gating or \gls{DVFS}. Due to these extended system modes, the PHY layer can achieve a better energy proportional behavior, compared to the fixed PHY layers that were used in prior works. With these extensions, the RA now has more selection choices, which extend the Pareto-frontier in terms of maximizing the achievable energy-efficiency gains.

\section{Receiver Energy Model} \label{subsec:em}
To perform the mode selection at run time according to \eqnref{eqn:EG}, the MAC layer of the client needs to estimate the power consumption of the modules composing the PHY layer for each possible system mode $\nu$. To enable such fine grained energy awareness, we model the energy consumption of the PHY layer by partitioning the contribution of each module based on their participation in the different phases of the transmission protocol shown in \figref{ch8:fig:frame}. 

\rvO{In order to understand the energy consumption of the PHY and MAC layers, a detailed energy model is needed.}
To this end, we have to understand the basic operating principles of the PHY layer under consideration. 
\rvO{In this paper, we use an IEEE 802.11n compliant PHY layer implementation based on \citep{Burg09} as a case study.}

\begin{figure}[t]
\centering
\includegraphics[width=1.0\columnwidth]{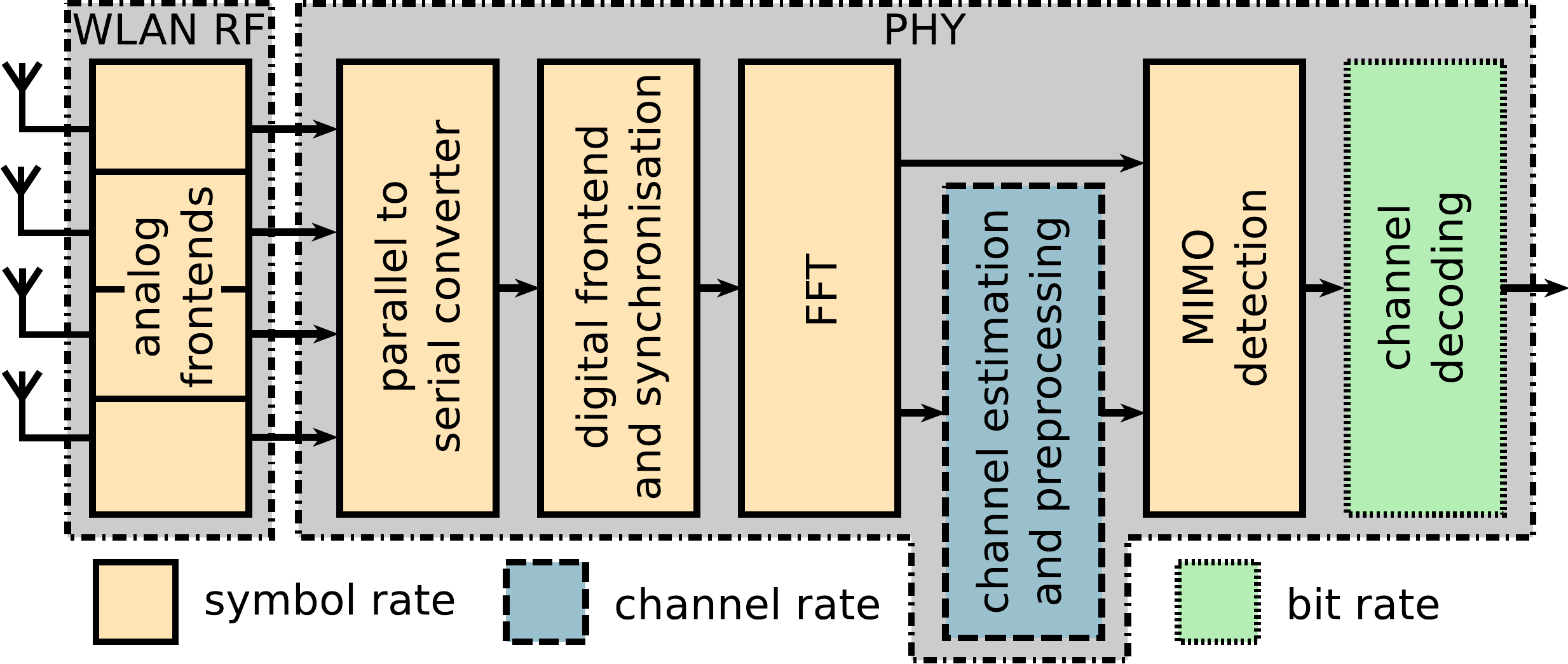}
\caption{Simplified PHY layer block diagram with the processing rate of the modules highlighted.}
\label{ch8:fig:rx}
\end{figure}

\subsection{Receiver Architecture}
A simplified architecture of a typical WLAN RF and PHY layer receiver is illustrated in \figref{ch8:fig:rx}. The signal from each of the \NRX\ antennas is fed into a separate analog/RF frontend where the radio frequency signal is down-converted to baseband and digitized. After a parallel to serial converter that time multiplexes the receive chains, a digital frontend is responsible for frame start detection as well as for \rvO{coarse-grained frequency synchronization. 
When a frame start is detected}, the signals are forwarded to a time-multiplexed \gls{FFT}. The frequency-domain representation of the training sequences of the received frame are fed to a channel estimation and preprocessing module. \rvT{There, the channel state information is extracted for each packet based on the training sequence and all computations for MIMO signal detection that only depend on the channel are performed. The output of the channel estimation \cite{haene2007,haene2008} and preprocessing \cite{luethi07a, senning10} module is fed into the subsequent MIMO detector together with the frequency-domain baseband samples of the data payload. The channel state information is further forwarded to the MAC layer as a basis for the RA for future data frames.} After MIMO detection, the bits (with or without reliability information) are forwarded to the channel decoding module that comprises a deinterleaver and a Viterbi decoder or an LDPC decoder.
 
In our transmission scenario, shown in \figref{ch8:fig:frame}, the analog/RF frontend is always turned on (in receive or transmit mode) except during sleep periods. The highest work load of the subsequent digital frontend is during the frame start detection and the frequency-offset estimation in the first training phase. The FFT processes data during the training phases, the header phase, and the data payload phase of the received frame. During the training phases the channel estimation and preprocessing module is also active. The subsequent MIMO detection and the channel decoding modules process both header and data payload.
 
The processing rate of the different modules is indicated in \figref{ch8:fig:rx}. Most modules run at the symbol rate of the transmission that is determined by the utilized signal bandwidth of the system. The channel estimation and preprocessing module has to process data during short periods once per frame, when the channel state information is estimated. Otherwise the module is idle. The channel decoding module processes at the coded bit rate of the transmission that varies in IEEE\,802.11n compliant systems over a large range from 13\,Mbps to 720\,Mbps, depending on the selected MCS.

\vspace*{-0.5cm}
\subsection{LUT based Energy Model}
We now consider the energy per bit of the receiver as our metric of main interest. To this end, we start by dividing the energy per received packet into three main contributions:
\begin{itemize}
\item A constant energy overhead for synchronization, header processing, training, channel-rate processing, and transmission of an ACK frame $e_{H}(\nu)$ in Joules. This part of the overall energy consumption depends partially on the choice of $\nu$ since different MCSs and different antenna configurations change also the duration of the training and have different energy costs.

\item The second contribution to the overall energy consumption comprises the RF and the baseband processing during the data phase. This part is characterized by the power consumption $p_{BB}(\nu)$ in Watts and the duration of the data phase. The latter is determined by the length $L$ and the throughput $\Phi(\nu)$ so that the corresponding contribution to the energy-per-frame amounts to $e_{BB}(\nu)=p_{BB}(\nu)\frac{L}{\Phi(\nu)}$.
 
\item The last contribution to the total energy consumption of the PHY layer at the receiver comprises mostly the channel coding which is typically carried out on a bit-by-bit basis and can be described by the energy per bit of the channel decoding $\eta_{CC}(\nu)$ and the length of the frame $L$ as $e_{CC}(\nu)=L\eta_{CC}(\nu)$.
\end{itemize}
 
Combining the three energy consumption contributions above and normalizing with the average number of successfully transmitted bits per packet, we obtain
\begin{equation}
 \eta(\nu)=\frac{\left(\frac{e_{H}(\nu)}{L}+\frac{p_{BB}(\nu)}{\Phi(\nu)}+\eta_{CC}(\nu)\right)}{1-P_e(\nu)}
\label{eq:energy_model}
\end{equation}
as our metric of interest for optimization of the energy spend per information bit at the PHY layer of the client.

One straight forward and effective implementation for the hardware characteristics of such an energy model can be realized with a \gls{LUT}, which can be configured off-line for any target design (such as the IEEE\,802.11n PHY layer that we use for the case study in this paper). \rvO{For such an energy model a detailed discussion of its parameters is provided in the following subsection.}

 \subsection{Energy Model Characterization}
In order to characterize the energy model based on the IEEE\,802.11n compliant PHY layer implementations presented in \citep{Burg09} and the analog frontend implementation discussed in \citep{Kumar13}, we will first elaborate on the components contributing to $e_{H}(\nu)$, $e_{BB}(\nu)$, and $e_{CC}(\nu)$. In a next step, we present an automated component wise calibration method that is based on post-layout gate-level power simulations of PHY layer implementations.

The first term of \eqnref{eq:energy_model}, the energy-overhead $e_{H}(\nu)$ is composed of
\begin{equation}
\begin{aligned}
e_{H}(\nu) & = e_{af}(\nu) + e_{df}(\nu) + e_{fft}(\nu) \\ 
         & \quad  + e_{chpp}(\nu) + e_{det}(\nu) + e_{ack},
\end{aligned}
\end{equation}
where $e_{af}(\nu)$, $e_{df}(\nu)$, $e_{fft}(\nu)$, $e_{chpp}(\nu)$, and $e_{det}(\nu)$ correspond to the energy used by the analog frontend, the digital frontend, the FFT, the channel estimation and the matrix preprocessing circuit, as well as of the detector, during processing of the overhead illustrated in \figref{ch8:fig:frame}. The final component $e_{ack}$ corresponds to the energy consumed by the PHY and RF to transmit an ACK frame to the AP.

To calibrate $e_{H}(\nu)$, we measured the average power consumption of the digital frontend, the FFT, the preprocessing module, the detector, and the RF. The average power consumption is then multiplied with the duration of the active time during the overhead processing of the corresponding components. While the active time during the header processing of the FFT, the preprocessing, and the detector loosely depend on the mode~$\nu$, the active time of sending an ACK frame is fixed. For a given MCS, we can assume, that the active time of each component contributing to the overhead processing energy is approximately constant. 

The second term of \eqnref{eq:energy_model}, the baseband energy consumption $p_{BB}(\nu)$, is composed of
\begin{align}
p_{BB}(\nu) = p_{af}(\nu) + p_{df}(\nu) + p_{fft}(\nu) + p_{det}(\nu),
\end{align}
where $p_{af}(\nu)$, $p_{df}(\nu)$, $p_{fft}(\nu)$, and $p_{det}(\nu)$ correspond to the power consumption of the RF, the digital frontend, the FFT, and the detector, respectively. 

The last term in \eqnref{eq:energy_model}, the energy-efficiency $\eta_{CC}(\nu)$ is only composed of the channel coding module. To calibrate the LUT for that module, we measure the energy consumption per bit of the channel decoding module.  

\begin{figure}
\begin{center}
\includegraphics[width=1\columnwidth]{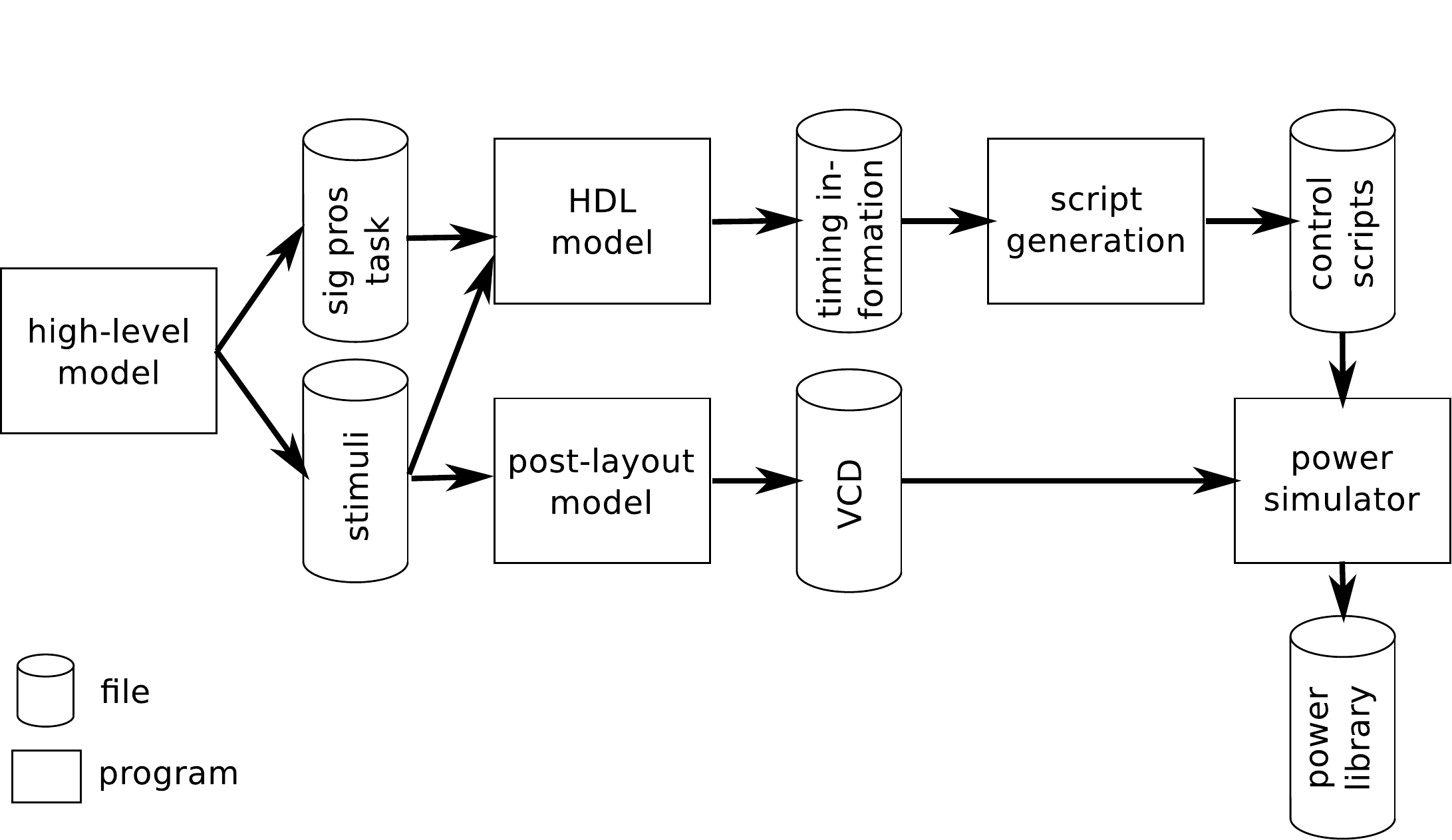}
\caption{Characterization of all modules in the PHY layer ASIC using an automated power library generation flow.}
\label{ch8:fig:genPwrFlow}
\end{center}
\end{figure}

The characterization flow for the energy model is illustrated in \figref{ch8:fig:genPwrFlow}. We first generate stimuli and then we monitor through simulations of the pre-synthesis hardware-description-language model the active modules for each system mode $\nu$
in order to accurately capture activity pattern of the actual hardware. The start and end time stamps of all active periods for each selected module are used to automatically generate the required scripts for post-layout power simulation. In parallel, the simulation of the post-layout circuit of the PHY layer implementation is performed. This simulation outputs \gls{VCD} files required for accurate power estimation. In a next step, the VCD files and the generated scripts are used to generate a library storing the power values for the selected modules during the periods of interest.
 
\subsection{Discussion on the Proposed Energy Model}
Before proceeding with the optimization of~\eqref{eq:energy_model} by properly choosing $\nu$ for a large, but fixed $L$, a brief discussion of the implications of this model and the dependency between its variables (through the choice of $\nu$) provides some insight into trends and ideas to simplify the estimation of the potential for energy savings. 
  
As a starting point for our considerations, we note that in the rather generic expression in \eqref{eq:energy_model}, $e_{H}(\nu)$, $p_{BB}(\nu)$, and $\eta_{CC}(\nu)$ all depend on $\nu$. However, in practice, we note at least for large $L$ the energy spent on overhead processing $e_{H}(\nu)$ is often insignificant\footnote{The payload phase of the transmission for low to medium throughput modes using a large $L$ is much longer \rvO{than} the duration of the header shown in \figref{ch8:fig:frame}.} compared to the other terms in \eqnref{eq:energy_model}. \rvO{Furthermore, we assume that for a traditional PHY implementation}, both $p_{BB}(\nu)$, and $\eta_{CC}(\nu)$ only have a limited dependency on the mode $\nu$. Hence, choosing $\nu$ to maximize $\Phi(\nu)$ trivially minimizes the \rvO{nominator} of \eqref{eq:energy_model} with diminishing returns due to the bias terms. Unfortunately, $P_e(\nu)$ also approaches one as $\Phi(\nu)$ increases which ultimately limits the achievable goodput as well as the energy-gains. To eliminate the rather complex relationship between $\nu$ and $P_e(\nu)$, we take advantage of the presence of a fast and conservative RA. For each channel realization we divide the available modes into two groups: one that is able to get a packet across with very high probability $P_e(\nu)\approx 0$ and a second group of modes that will almost surely fail ($P_e(\nu)\approx 1$). Under these assumptions, the choice of the highest-rate mode that is still reliable is clearly the most favorite strategy and little potential would exist for further energy-efficiency optimization. 

However, if different modes are able to provide significantly different $e_{H}(\nu)$, $p_{BB}(\nu)$, or $\eta_{CC}(\nu)$ for the same channel realization, while still maintaining $P_e(\nu) \approx 0$, we may be able to obtain further energy-efficiency gains over the \rvO{straightforward} choice for the mode that provides the best goodput.
 
\section{System Mode Extension for Enhanced Energy-Efficiency}\label{sec:phyMod}
To enable further energy-efficiency improvements beyond the ones associated with the natural choice of the highest-rate mode, the baseline PHY layer implementation must first be improved. The main idea is to first introduce a more energy proportional behavior in a sense that $e_{H}(\nu)$, $p_{BB}(\nu)$, and $\eta_{CC}(\nu)$ scales better with the different processing requirements of the different modes~$\nu$. \rvO{In the second step,}
we then introduce new modes which are not necessarily optimal in terms of their goodput, but may still provide an overall energy-efficiency advantage by further reducing processing requirements at the expense of rate or reliability (error rate).

\subsection{Energy Proportionality Without Impact on Goodput Through DVFS}\label{subsec:noPerf}
A well designed physical layer implementation already provides a certain degree of energy-proportionality. Hence, the energy consumption of such a PHY depends more or less linearly on the number of operations required to extract the payload data from the received frame. A good example is the channel decoding, which contributes to \eqref{eq:energy_model} a constant energy per decoded bit. Unfortunately, such circuits do not exploit differences in the time available to process each individual bit. To take advantage of different throughput requirements in different modes, we notice that in very large integrated circuits the maximum operating frequency of a digital circuit scales approximately linearly with the supply voltage (within reasonable range). However, the power consumption of the circuit scales quadratically with the supply voltage. Therefore, we can translate relaxed throughput requirements (in terms of operations per second) into further energy-savings per operation, an idea that is commonly referred to as \gls{DVFS}.

Practical applications of DVFS \cite{Larsson11} are clearly associated with many difficulties and with overhead (e.g., voltage regulators and alike). \rvT{In~\cite{Villar13} it is shown that integrated voltage regulators with a conversion-efficiency better than 80--90\% and a reasonable circuit size are feasible in modern CMOS technology.
More specifically, integrated DC-DC converters providing an output current with several amps per square millimeter silicon area have been proposed in \citep{chang2010} and \citep{Andersen2014}. Hence voltage regulators for circuits consuming few hundred milliwatt require only a small fraction of a square millimeter silicon area.  
However, in order to explore the limits of energy-optimal data transmission, regardless of the technical issues on circuit level we intentionally neglect the associated overhead for voltage conversion in this paper.}
 
\begin{figure}[t]
\includegraphics[width=\columnwidth]{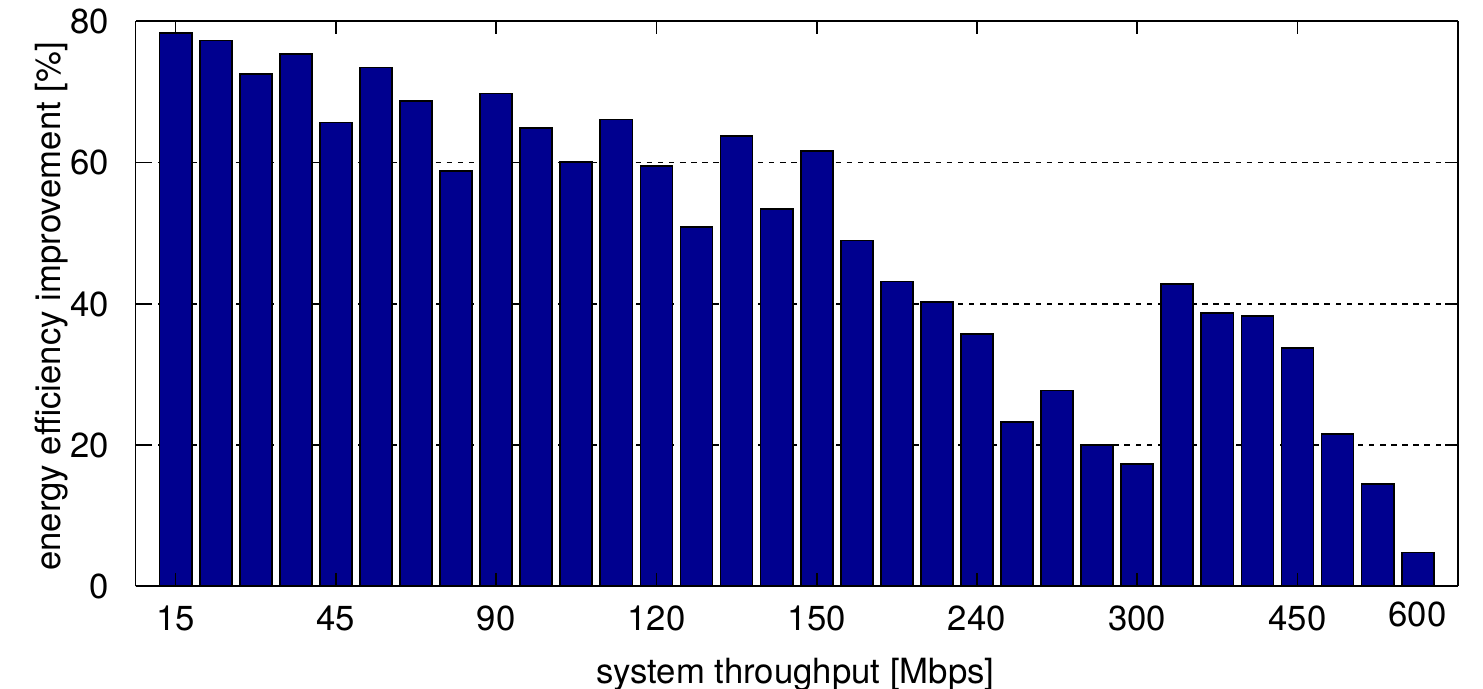}
\caption{Relative energy-efficiency improvement of the channel coding
  module using DVFS for all supported transmission schemes.}
\label{ch8:fig:ccRelImp}
\end{figure}
 
\paragraph*{Application of DVFS to channel decoding:} 
\ A first obvious opportunity to apply DVFS in a conventional IEEE\,802.11n compliant receiver is the channel decoding module. For this module, the processing rate varies significantly (from 13\,Mbps to 720\,Mbps). For transmission modes with a system data rate up-to 300\,Mbps, one decoding core is used. On the other hand, two cores are used for larger data rates. With DVFS, this part of the receiver can take advantage of the reduced rate and \rvO{provide better overall}
energy-per-bit without negative impact on error rate performance of the receiver. 

The relative energy savings for the channel decoding module applying DVFS are shown in \figref{ch8:fig:ccRelImp}. It can be seen, that the lower the data rate, the better the energy savings from DVFS as the clock frequency and consequently the supply voltage can be reduced. Further, it can be seen, when the number of channel decoding cores changes from one to two, as specified in the IEEE\,802.11n standard, then the achievable energy-efficiency gain is again increased, as the data rate per decoder core is reduced.

\subsection{Energy Proportionality With Impact on Goodput}
Two other possibilities for enhancing the energy proportional behavior of the PHY layer, which can affect the goodput are the following:

\paragraph*{Sub-optimal algorithms:}
\ A possible modification to improve energy proportional behavior is to take advantage of the fact that in some situations different receiver algorithms with noticeable complexity difference exhibit a similar error-rate behavior. Hence, choosing an algorithm with higher computational-complexity may still not allow for a goodput improvement that out-weights the higher energy-cost.
 
A good example is the choice of the MIMO detector, where a solution with close to \gls{ML} or \gls{APP} performance can be combined with a low-complexity MIMO detector. The ML or APP detector provides good performance required for bad conditions, (e.g., low \gls{SNR} or ill-conditioned MIMO channels), while the low-complexity alternative uses less energy. For conditions where the occurrence of a frame error is independent of the detector (e.g., very high SNR, very low SNR), the low-complexity alternative achieves obviously better energy-efficiency.

\paragraph*{Exploiting active antenna selection for applying DVFS:} 
A second opportunity to apply DVFS exists in case of reduced-complexity receiver configurations that may be suboptimal in terms of error rate performance but can be advantageous in terms of energy consumed. Since the corresponding reduced complexity modes will generally only allow for a goodput that is lower than that provided by more complex modes, they can only be advantageous when savings in $e_{H}(\nu)$, $p_{BB}(\nu)$, and $\eta_{CC}(\nu)$ make up for the goodput loss. The most obvious target for such additional modes is the choice of the number of active antennas \NRX\ at the receiver~\cite{Sanayei04}. Such adaptation allows to roughly linearly scale the power of the RF frontend with \NRX, but also limits the number of spatial streams to \NSS<\NRX. The baseband processing can take further advantage of this scaling due to the reduced number of operations (e.g., number of multiplications and additions) but also benefits from DVFS since fewer receive chains must be processed in a time-interleaved fashion and therefore more time is available per active receive chain.

\section{Evaluation} \label{sec:simRes}
In this section we explore the limits of the gains in energy-efficiency that can be achieved with the proposed modifications to the PHY layer combined with the proposed energy-guided RA scheme. Furthermore, we compare the proposed RA schemes with the classical GG RA. We perform this evaluation on an IEEE\,802.11n compliant wireless communication system for two different scenarios described in the following paragraphs.

\subsection{Scenarios}
In our evaluation, we consider two different scenarios: In the first scenario, the AP transmits frames with a fixed length $L$ of 1.5\,kB over a Rayleigh block-fading channel. For this simple scenario we compare the impact of all three discussed RA schemes in detail. In the second scenario the AP transmits frames with a varying length. To this end, we enable the AP to perform aggregation\footnote{Frame error rate remains independent of the number of aggregated frames (individual ACK in the same ACK frame). Still, overhead (PHY and ACK for small packets individually) can be avoided, which motivates aggregation of multiple frames.} of up to 16 frames, each with 1.5\,kB.

For both scenarios, we considered the RF given in \citep{Kumar13} combined with the IEEE\,802.11n compliant PHY layer given in \citep{Burg09}, with up to 4 spatial streams and up to 4 receive antennas using transmissions with 40\,MHz bandwidth. For channel coding we use only the convolutional code defined in the standard. The 32 mandatory MCSs defined in IEEE\,802.11n result  in a variable frame duration for a given $L$.
 
For $10^3$ channel realizations, the reception of a frame for all considered system modes have been simulated. In particular the simulated system modes include all mandatory MCSs of the IEEE\,802.11n standard. For each MCS, the number of active receive antennas varies between \NSS\ (the minimal number of receive antennas required for the specific MCS) and 4. In addition, all MCSs have been simulated with a hard-output lattice reduction aided linear \gls{MMSE} MIMO detector (LRALD) \cite{senning14}, and a low-complexity soft-output MMSE detector, resulting in 112 different system modes\footnote{32 transmission modes received with 4 receive chains without applying DVFS; 32 transmission modes with 4 receive chains using DVFS at the channel decoder; 24 transmission modes using 3 receive chains and DVFS wherever applicable; 16 transmission modes using 2 receive chains and DVFS wherever applicable; 8 transmission modes using 1 receive chain and DVFS wherever applicable.} for each MIMO detector algorithm comprising $\Omega_{EG}$.



\subsection{Ideal Estimation of Channel Conditions}
As we discussed in \secref{sec:propapp}, any RA requires an accurate estimation of $P_e(\nu)$ for the actual channel conditions. Since the purpose of this paper is to explore the potential and the limits of energy-awareness we intentionally assume a genie-aided approach for the RA. This simplification avoids introducing uncertainties due to specific RA strategies and it simplifies the full explanation of the setup under consideration. The genie provides an upper bound on goodput and energy-efficiency by perfectly predicting transmission failures by sending each packet in all modes for each channel and noise realization which -- of course -- is only feasible in simulation but could be approached in practice with carefully designed algorithms. Hence, $P_e(\nu)$ is either one or zero which boils down to limiting the selection to the error-free modes according to a specific optimization criterion\footnote{Practical implementation to estimate $P_e(\nu)$ can be found in \cite{biaz08}.}.

\begin{figure}
 \centering
 \includegraphics[width=0.9\columnwidth]{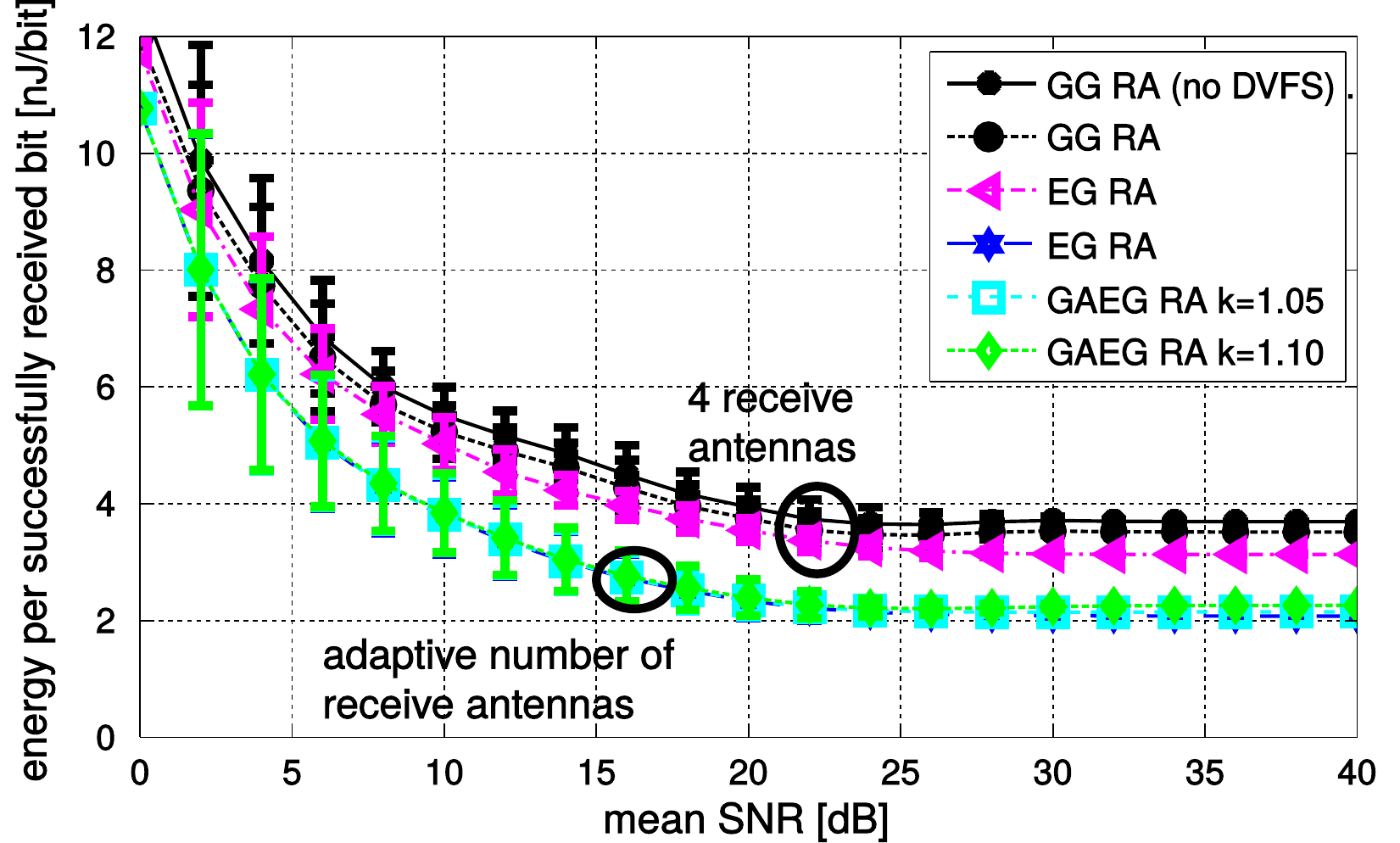}
 \caption{Energy consumption per successfully transmitted bit.}
 \label{fig:absEeff}
\end{figure}

\begin{figure}
 \centering
 \includegraphics[width=0.9\columnwidth]{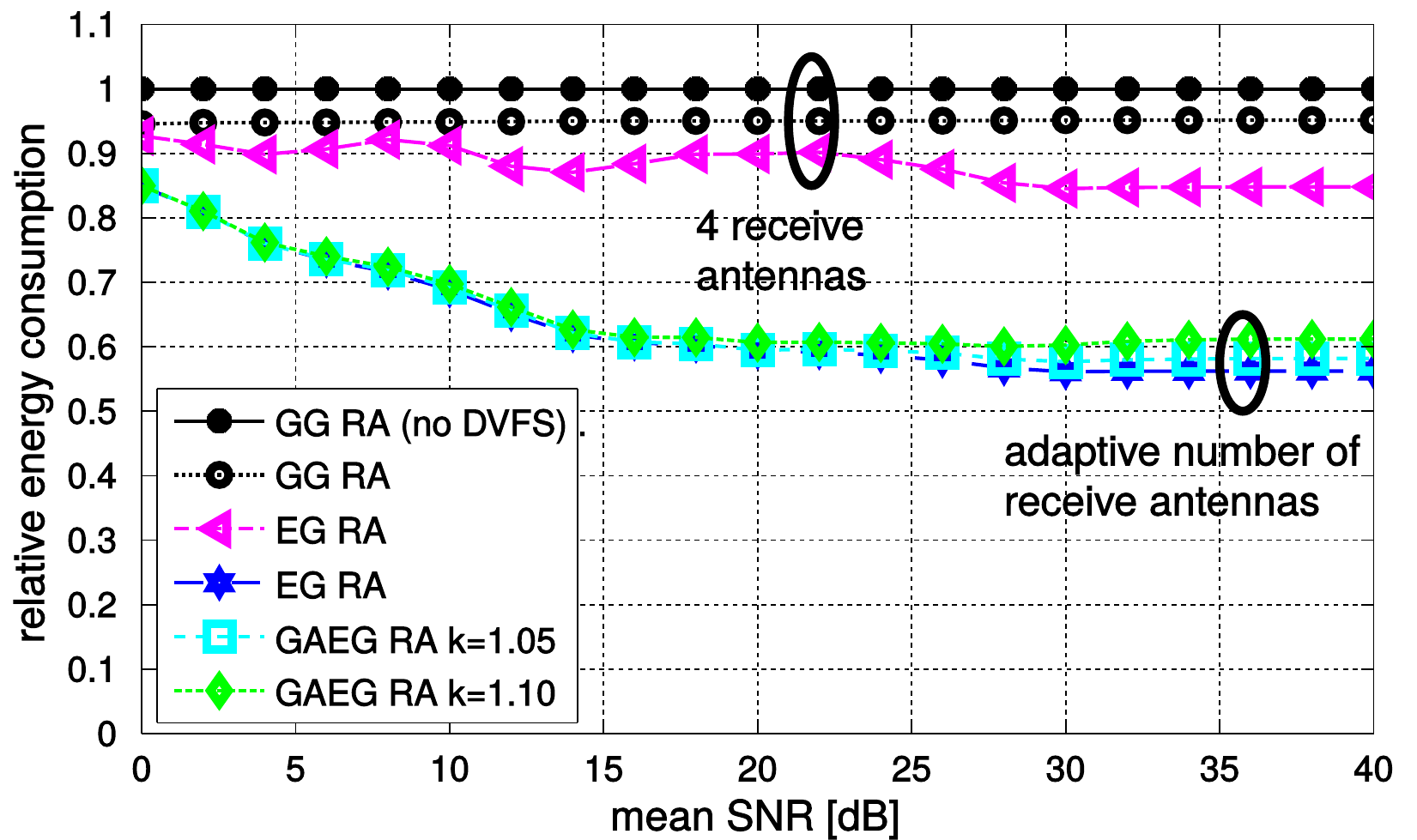}
 \caption{Relative power consumption with respect to GG RA without DVFS.}
 \label{fig:relEff}
\end{figure}

\begin{figure}
 \centering
 \includegraphics[width=0.9\columnwidth]{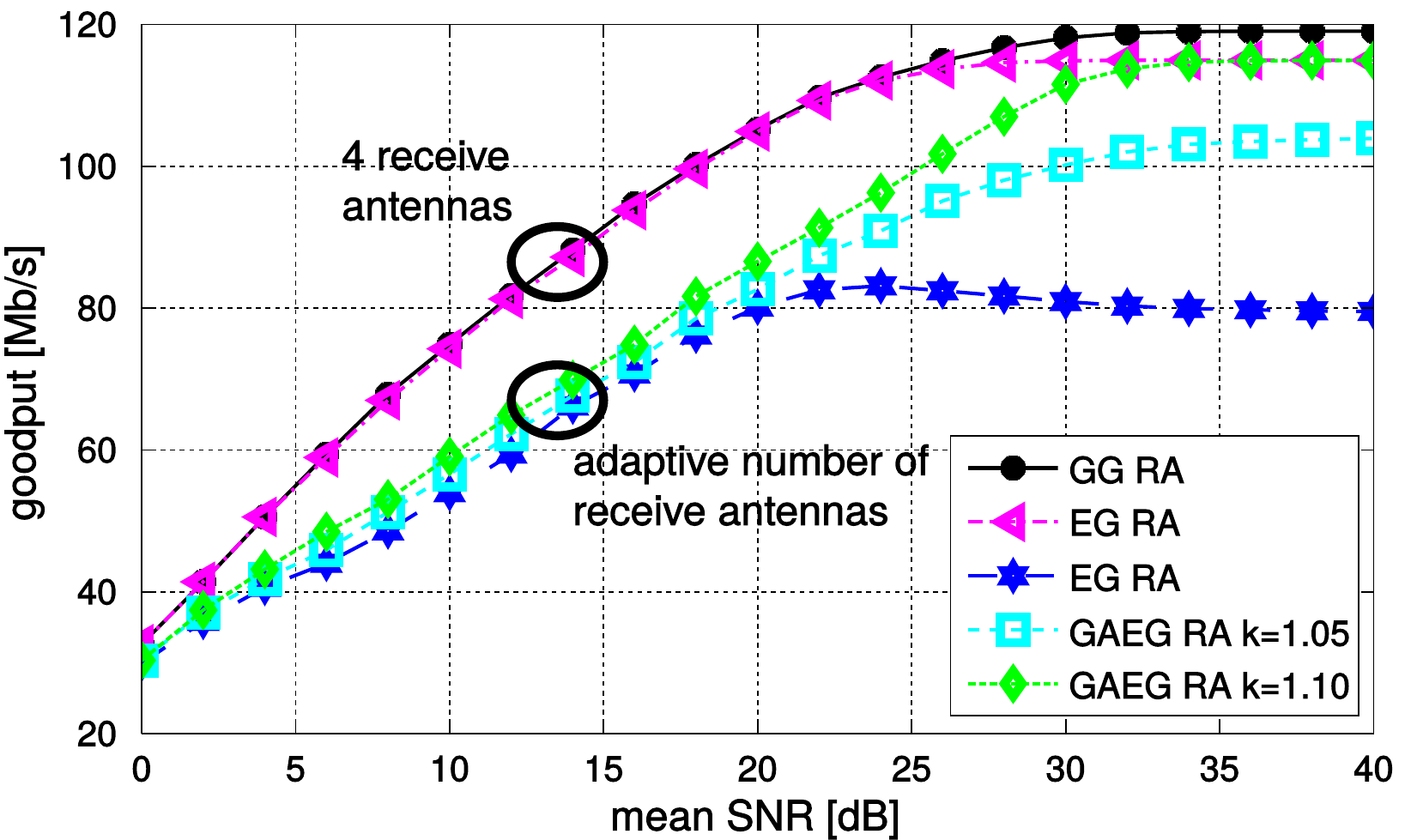}
 \caption{Throughput achieved by the different schemes.}
 \label{fig:absTh}
\end{figure}

\subsection{Results for a Fixed Frame Length} \label{ch8:sec:resFix}
In \figref{fig:absEeff} to \figref{fig:absTh} we compare the impact of the  proposed modifications in conjunction with the considered RA strategies on absolute and relative energy per bit, as well as on goodput.
 
We compare all obtained curves to the reference one corresponding to a GG RA without the PHY layer modifications proposed in \secref{sec:phyMod}. This baseline implementation applies no DVFS and always employs 4 receive chains resulting in the worst energy consumption per successfully received bit as shown in \figref{fig:absEeff}. However, such GG approach leads to the highest goodput over the entire SNR range as shown in \figref{fig:absTh}.

\begin{itemize}
\item By assuming only PHY modifications, under a GG RA, no significant energy-efficiency gains can be achieved over the baseline implementation. In particular, GG RA never sacrifices error rate performance by reducing the number of active antennas or by using a sub-optimal MIMO detector. Hence, the only PHY modification that can be applied is DVFS in the channel coding module. Since channel decoding is only a small fraction of the overall receiver power consumption, we obtain only the marginal energy reduction of 5\% as shown in~\figref{fig:relEff}.
 
\item By modifying the RA to EG RA, while still applying DVFS only in the channel decoding, without the capability to scale the number of active receive chains, the energy per bit is reduced by up to 15\% compared to the baseline case. Although the EG RA improves the energy-efficiency compared to the baseline implementation and the previous considered case, the restriction to exactly 4 active receive antennas does not reveal the maximum gains. Note that, the goodput does not deteriorate significantly compared to the baseline implementation.

\item The maximum average efficiency gains up to 44\% are obtained  in case of joint application of all proposed PHY layer modifications under an EG RA (DVFS applied in the channel coding module along with DVFS based on active receive chain selection, as well as selection of an appropriate MIMO detector for the specific channel condition). However, such savings come at the cost of a reduced goodput, as shown in \figref{fig:absTh}. In any case, most applications such as today's compressed HD videos on popular video-on-demand platforms or high quality audio streams require a much lower data rate than the one achieved by our approach. Hence, the proposed approach can still satisfy the data rate requirements of most popular applications while significantly improving the energy-efficiency of the battery operated device.
 
\item For applications requiring much higher goodput and in case that minor energy-efficiency sacrifice is acceptable, then our results show that GAEG RA can provide a reasonable compromise between EG RA and GG RA. \rvT{As shown in \figref{fig:absEeff} and \figref{fig:relEff} GAEG RA with a factor $k=1.05$ or $k=1.10$ reduces the efficiency only slightly compared to the best achievable gain in case of EG RA.} However, in the high SNR region (i.e., the region where high data rates can be achieved) the maximum achievable goodput is restored to 97\% compared to the goodput achieved by the GG RA scheme, as shown in \figref{fig:absTh}.
\end{itemize}


In order to show the efficacy of the EG RA, \tblref{tbl:expRes2} and \tblref{tbl:expRes3} demonstrate the percentaged distribution of the selected modes and the corresponding energy consumption per successfully received bit for the different SNR values. In case of 15\,dB SNR, the EG RA chooses most of the time the linear MMSE MIMO detector, because the LRALD has a higher computational complexity and therefore is more power hungry, but rarely improves the reception of the frames under such channel conditions. However in the high-SNR regime, where ill-conditioned channels have a more prominent impact, the choice of LRALD is still preferable for the EG RA in many cases, as shown in \tblref{tbl:expRes3}.


\subsection{Results for Varying Frame Lengths} \label{ch8:sec:resVar}
\begin{figure}[t]
\begin{center}
\includegraphics[width=1\columnwidth]{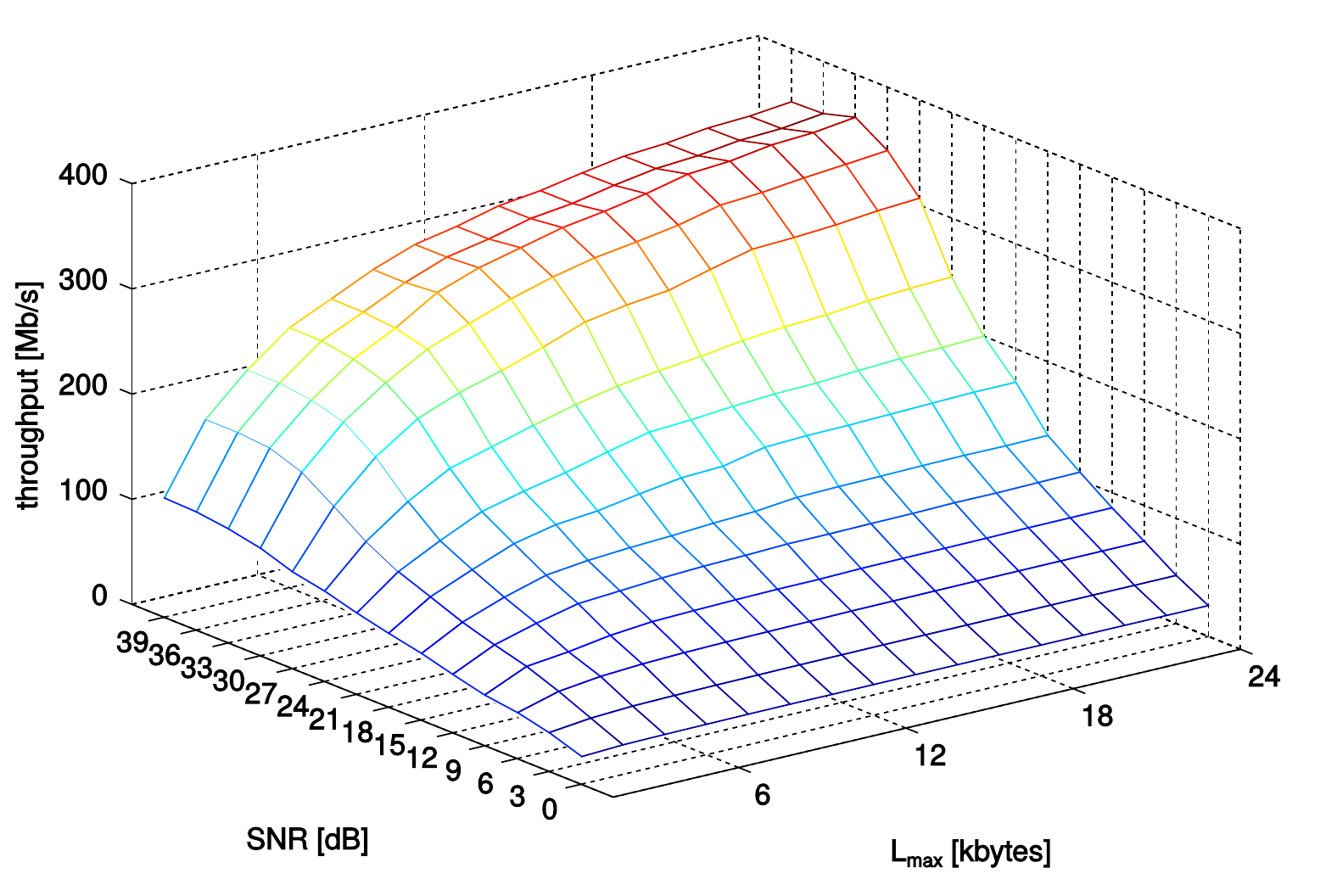}
\caption{Throughput of the proposed optimized system.}
\label{fig:meshTh}
\end{center}
\end{figure}

In order to show the relative energy improvement for the overall battery operated device in an office environment, we consider the case of aggregated frames with the IEEE TGnC~\cite{TGN04} channel model, as described in the IEEE~802.11n standard. To this end, we evaluate for varying frame sizes in \figref{fig:meshEff} to \figref{fig:meshTh} the achieved energy-efficiency and goodput of the proposed PHY layer modifications under an EG RA scheme, compared with the ones achieved by the baseline implementation. For this setup, we allow the AP to aggregate up to 16 frames. Both, the GG RA and the EG RA are allowed to select the number of aggregated frames for each transmission freely such that the resulting total frame length $L$ is between 1.5\,kB and $L_{max}\leq24$\,kB, while each of the aggregated frames is acknowledged independently in a single ACK frame.

The resulting relative energy consumption with respect to the SNR and the maximum frame length $L_{max}$ (which is an integer multiple of 1.5\,kB) is depicted in \figref{fig:meshEff}. It can be observed that the best relative energy consumption is achieved for short frames and moderate to high SNR reaching up to 36.6\%. In this moderate to high SNR region the RA can take advantage of many 

\begin{minipage}[l]{\columnwidth}
%
\captionof{table}{Experimental results at 15\,dB SNR}
\label{tbl:expRes2}
\vspace{0.5em}
\setlength{\tabcolsep}{3pt}
\resizebox{0.94\columnwidth}{!}{
\begin{tabular}{ccr@{.}l|cccr@{.}l}
\toprule[0.15em]
\multicolumn{4}{c|}{GG RA} & \multicolumn{5}{c}{EG RA (var. ant.)}\\
MCS    & [nJ/bit] & \multicolumn{2}{c|}{\%}  & MCS & alg.      &  [nJ/bit] & \multicolumn{2}{c}{\%}\\\midrule
12     & 4.21     &  65&0                    & 12  & MMSE         & 3.85      &  0&5 \\            
13     & 3.58     &   6&8                    & 3   & MMSE         & 3.27      & 14&6 \\            
14     & 3.46     &   0&5                    & 4   & MMSE         & 2.49      & 22&3 \\            
18     & 5.13     &   1&0                    & 5   & MMSE         & 2.07      &  4&9 \\            
19     & 4.39     &   2&4                    & 6   & MMSE         & 1.94      &  1&0 \\            
20     & 3.58     &  16&0                    & 10  & MMSE         & 3.44      &  1&5 \\            
27     & 3.88     &   8&3                    & 11  & MMSE         & 2.81      & 31&1 \\            
       &          & \multicolumn{2}{c|}{\ }  & 12  & MMSE         & 2.23      &  2&7 \\            
       &          & \multicolumn{2}{c|}{\ }  & 18  & MMSE         & 3.38      &  8&2 \\            
       &          & \multicolumn{2}{c|}{\ }  & 19  & MMSE         & 2.88      &  8&2 \\            
       &          & \multicolumn{2}{c|}{\ }  & 26  & MMSE         & 3.62      &  1&0 \\
       &          & \multicolumn{2}{c|}{\ }  & 27  & MMSE         & 3.09      &  2&9 \\
       &          & \multicolumn{2}{c|}{\ }  & 14  & LRALD        & 3.56      &  1&0 \\\midrule
\multicolumn{4}{c|}{average: 4.05\,nJ/bit}           & \multicolumn{5}{c}{average: 2.83\,nJ/bit} \\
       
\bottomrule[0.15em]
\end{tabular}}
\vspace*{0.5cm}

\captionof{table}{Experimental results at 30\,dB SNR}
\label{tbl:expRes3}
\vspace{0.5em}
{
\setlength{\tabcolsep}{3pt}
\resizebox{0.94\columnwidth}{!}{
\begin{tabular}{ccr@{.}l|cccr@{.}l}
\toprule[0.15em]
\multicolumn{4}{c|}{GG RA} & \multicolumn{5}{c}{EG RA (var. ant.)}\\
MCS &  [nJ/bit] & \multicolumn{2}{c|}{\%}  & MCS & alg.  & [nJ/bit] & \multicolumn{2}{c}{\%}\\\midrule
                    22    & 3.07     & 1&0                      &  8  & MMSE         & 1.85     & 7&3     \\            
                    23    & 2.92     & 71&8                     & 14  & MMSE         & 1.93     & 1&0     \\            
                    29    & 2.90     & 0&5                      & 15  & MMSE         & 1.89     & 3&9     \\            
                    31    & 2.80     & 26&7                     & 16  & MMSE         & 1.85     & 7&3     \\            
                          &          & \multicolumn{2}{c|}{\ }  & 23  & MMSE         & 1.92     & 0&5     \\            
                          &          & \multicolumn{2}{c|}{\ }  & 24  & MMSE         & 1.82     & 48&5    \\            
                          &          & \multicolumn{2}{c|}{\ }  & 4   & LRALD        & 1.85     & 12&1    \\            
                          &          & \multicolumn{2}{c|}{\ }  & 11  & LRALD        & 1.89     & 1&9     \\            
                          &          & \multicolumn{2}{c|}{\ }  & 12  & LRALD        & 1.82     & 17&0    \\            
                          &          & \multicolumn{2}{c|}{\ }  & 20  & LRALD        & 1.90     & 0&5     \\\midrule
\multicolumn{4}{c|}{average: 2.89\,nJ/bit}           & \multicolumn{5}{c}{average: 1.83\,nJ/bit} \\
       
\bottomrule[0.15em]
\end{tabular}}
}
\vspace*{0.5cm}

\includegraphics[width=1\columnwidth]{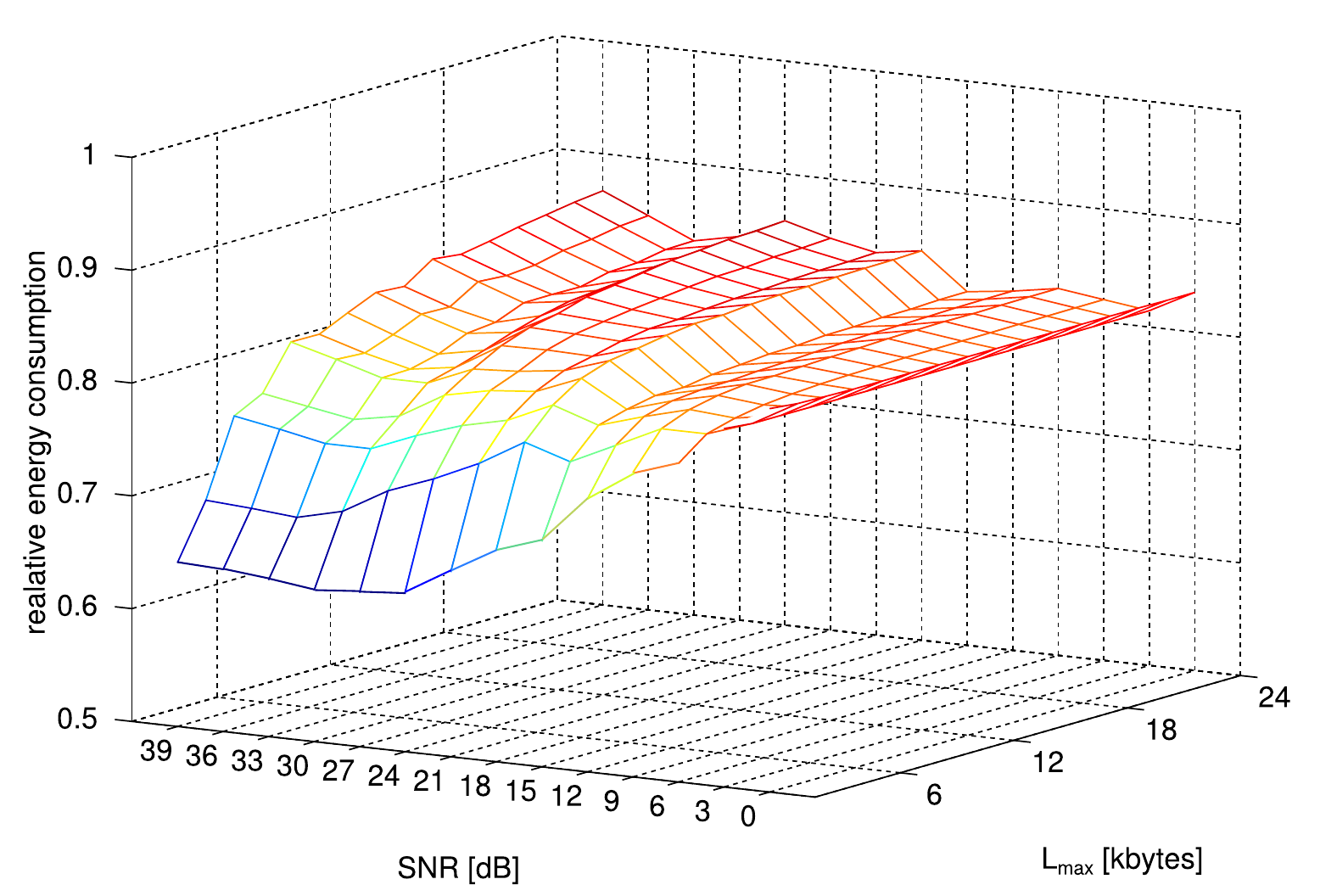}
\captionof{figure}{Relative energy-efficiency for different maximum frame lengths and different channel quality.}
\label{fig:meshEff}
\end{minipage}

\hspace*{-0.5cm}operation modes with a $P_e(v)\approx 0$. It is worth mentioning that the relative energy consumption of the proposed system is in all regions at least 13.3\% better than that of the baseline implementation. This indicates that indeed our method can improve the energy-efficiency of mobile clients for realistic channel models.

Finally, we show the impact of the EG RA on the achieved goodput in \figref{fig:meshTh}. While we expected a reduced goodput for non-aggregated frames, our data shows that EG RA achieves for aggregated frames a goodput close to the goodput achievable by GG RA. Therefore there is no need for a GAEG RA and any compromise in terms of energy-efficiency can be avoided.

\section{Conclusion}\label{sec:conc}
In this paper, we depart from conventional goodput optimized communication systems and present a cross-layer approach that involves various design, analysis and run-time techniques for improving the energy-efficiency of packet based communication systems. The proposed approach considers all blocks within the PHY layer of battery operated devices that participate in the communication link and optimizes the energy-efficiency by jointly considering the system level and hardware level parameters within the MAC and PHY layer.

The energy efficiency optimization is achieved with the introduction of an energy-guided rate adaptation scheme which is combined with several modifications of the PHY layer to enable maximum energy savings.
Several new modes of operation with reduced receiver complexity are added to further improve the energy proportional behavior of the battery operated device. Fine grained energy models of PHY are developed for propagating such behavior to the MAC layer and allowing the energy-guided rate adaptation to select the optimum system mode under any given condition and requirement. 

The numerical results indicate that our approach can achieve up-to 44\% energy efficiency improvement in an IEEE\,802.11n system. This is achieved by properly selecting the system mode along with the right degree of voltage and frequency scaling and the selection of the appropriate algorithm for data processing when possible. We further show that a goodput-aware energy-guided rate adaptation can provide a reasonable compromise between energy and goodput if the goodput achieved by the energy-guided rate adaptation is insufficient. In any case, our work, encourages a paradigm shift towards an energy-guided RA with a properly energy proportional PHY layer, thus motivating  further work towards this direction. 

\begin{acknowledgements}
This work was partially supported by the Hasler Foundation project WiLANCE and the EU Marie Curie DARE grant 304186. 
\end{acknowledgements}


\bibliographystyle{abbrv}

\end{document}